# ROSSLER NONLINEAR DYNAMICAL MACHINE FOR CRYPTOGRAPHY APPLICATIONS

**Sunil Pandey**[*]
M.Tech. (Info Sec)
Computer Sc. & Engg.
MANIT Bhopal 462 051

**Praveen Kaushik**
Lecturer
Computer Sc. & Engg.
MANIT Bhopal 462 051

**Dr. S.C. Shrivastava**
Professor & Head
Elec & Comm Engg.
MANIT Bhopal 462 051

[*]Corresponding Author
e-mail: algosunil@yahoo.co.in

## ABSTRACT

In many of the cryptography applications like password or IP address encryption schemes, symmetric cryptography is useful. In these relatively simpler applications of cryptography, asymmetric cryptography is difficult to justify on account of the computational and implementation complexities associated with asymmetric cryptography. Symmetric schemes make use of a single shared key known only between the two communicating hosts. This shared key is used both for the encryption as well as the decryption of data. This key has to be small in size besides being a subset of a potentially large keyspace making it convenient for the communicating hosts while at the same time making cryptanalysis difficult for the potential attackers. In the present work, an abstract Rossler nonlinear dynamical machine has been described first. The Rossler system exhibits chaotic dynamics for certain values of system parameters and initial conditions. The chaotic dynamics of the Rossler system with its apparently erratic and irregular characteristics and extreme sensitivity to the initial conditions has been used for the design of the cryptographic key in an attempt to increase the confusion and the challenge for the potential attackers.

## Summary of Results

In this paper symmetric cryptography schemes for ensuring data confidentiality and integrity in simple cryptography applications based on the concept of a Rossler nonlinear dynamical machine has been described.

## INTRODUCTION

Simple systems like for example password encryption or IP address encryption generally use symmetric cryptography. For such systems standard asymmetric cryptography techniques like RSA or Elliptic Curve Cryptography or symmetric cryptography techniques like DES or AES are overkill.

### Data Confidentiality

Data that is to be kept confidential includes passwords and IP addresses. The data is encrypted with a secret key that only intended receivers possess thereby achieving confidentiality. The Rossler system is conceptually simple and is not based on the concept of rounds, which are computation intensive as well as complex and could therefore find application here.

### Data Integrity

Even though an adversary may not be able to steal information it may be possible to change data during transmission. Data integrity ensures that data has not been altered in transit.

**Chaotic Systems**

Chaos theory investigates the dynamics of nonlinear systems with relatively few sub-units or degrees of freedom that exhibit random-like, yet perfectly deterministic dynamical behavior and a high sensitivity to initial conditions. According to the Poincare-Bendixson theorem, continuous time nonlinear dynamical systems with at least three degrees of freedom can exhibit chaotic dynamics. Properties of continuous time chaotic systems useful in the design of cryptographic algorithms include the apparently erratic but perfectly deterministic dynamics of such systems, the high-sensitivity of such systems to infinitesimal changes in initial conditions, computational irreducibility of system trajectories, real phase space, and real system parameters. Digital realization of chaotic dynamics is through non integer arithmetic which is an approximation to continuous-value systems.

**Rossler System**

The Rossler system has been studied by Otto Rossler and arose from his work in the field of chemical kinetics. The Rossler system is described by a system of 3 coupled nonlinear ordinary differential equations:

$$dx / dt = - y - z$$
$$dy / dt = x + a y$$
$$dz / dt = b + z ( x - c )$$

where a = 0.2, b = 0.2, c = 5.7

This series also does not form limit cycles or reach a steady state. Like other chaotic systems the Rossler system is sensitive to the initial conditions, and two close initial states will diverge, with increasing number of iterations.

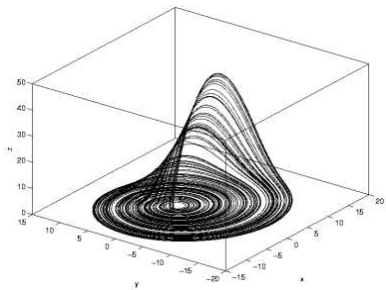

Fig 1: 3-D phase portrait of the Rossler attractor

**Integration Solver**

The Runge-Kutta method for a system of ordinary differential equations is explained here. Suppose there are *m* variables $x_1, x_2, ... x_m$ each of which vary over time. Suppose further there are *m* differential equations for these *m* variables given by:

$$x_1' = f_1(x_1, x_2, ..., x_m)$$
$$x_2' = f_2(x_1, x_2, ..., x_m)$$
$$...$$
$$x_m' = f_m(x_1, x_2, ..., x_m)$$

There are no derivatives on the right hand side of any of those equations, and there are only first derivatives on the left hand side. These equations can be summarized in vector notation as $\mathbf{x}' = \mathbf{f}(\mathbf{x})$ where $\mathbf{x} = (x_1, x_2, ..., x_m)$ and we use the "vector of functions" concept where $\mathbf{f} = (f_1, f_2, ..., f_m)$. Next we label our time states $\mathbf{x}_n$, $\mathbf{x}_{n+1}$ which are separated by time interval of length $h$. That is, $\mathbf{x}_n$ is the value of the $m$ variables at time $t_n$. And $x_{1,n}$ is the value of the first variable $x_1$ at time $t_n$.

$$\mathbf{x}_n = (x_{1,n}, x_{2,n}, ..., x_{m,n})$$
$$\mathbf{x}_{n+1} = (x_{1,n+1}, x_{2,n+1}, ..., x_{m,n+1})$$

Suppose we have the state of the simulation at time $t_n$ as $\mathbf{x}_n$. To compute the state a short time $h$ later and put the results into $\mathbf{x}_{n+1}$, the Runge-Kutta method does the following:

$$\mathbf{a}_n = \mathbf{f}(\mathbf{x}_n)$$
$$\mathbf{b}_n = \mathbf{f}(\mathbf{x}_n + {}^h\!/_2\, \mathbf{a}_n)$$
$$\mathbf{c}_n = \mathbf{f}(\mathbf{x}_n + {}^h\!/_2\, \mathbf{b}_n)$$
$$\mathbf{d}_n = \mathbf{f}(\mathbf{x}_n + h\, \mathbf{c}_n)$$
$$\mathbf{x}_{n+1} = \mathbf{x}_n + {}^h\!/_6\, (\mathbf{a}_n + 2\, \mathbf{b}_n + 2\, \mathbf{c}_n + \mathbf{d}_n)$$

The new vector $\mathbf{x}_{n+1}$ gives you the state of the simulation after the small time h has elapsed.

**Rossler Machine**

| System Parameter Vector, $\mathbf{v}_{sp}=[x\ y\ z]^T$ | Integration Solver (RK 4) | Output State Vector |
|---|---|---|
| Initial State Vector, $\mathbf{v}_{is}=[a\ b\ c]^T$ | | |
| Number of Time Steps, N | | |

**Fig 3: Rossler Machine**

The Rossler machine shown in Fig 3 computes the dynamics of the Rossler system. This machine takes as inputs the system parameter vector whose components are the three system parameters, the initial state vector whose components are the initial states of the three variables and the scalar number of iterations or time steps of the integration solver. The three components of the system parameter vector and the initial state vector are floating point numbers while the number of time steps is a positive integer. The output of the Rossler machine is the state vector, i.e., the values of the state variables at the end of the pre-specified number of time steps.

The Rossler machine can be realized in hardware.

**Related Work**

References mentioned in [1], [2], and [4] through [8] specifically deal with encryption and decryption schemes based on chaos theory.

# Results

## Simulation

Simulation has been performed in Scilab environment. Figure 2 shows the plot of the dynamics of the first dependent variable of the Rossler nonlinear dynamical system. Simulation has been performed for $\mathbf{v}_{sp} = [0.2 \ 0.2 \ 5.7]^T$, $\mathbf{v}_{is} = [0.0001 \ 0.0001 \ 0.0001]^T$, N=500 and step size parameter in the Runge Kutta solver is 0.1

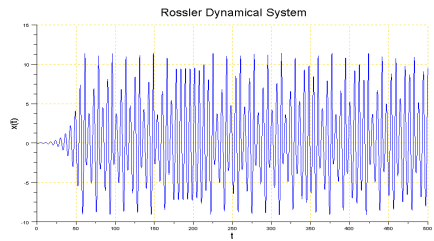

Fig. 2: Dynamics of 1$^{st}$ Rossler Variable

## Data Confidentiality Scheme

Individual characters of the plaintext message are first mapped to real numbers. These values are the initial states of any one variable, say x. The Rossler machine is run with given state parameter and initial state vectors and number of time steps. Values at the end of iterations are taken as the ciphertext corresponding to the individual character. Ciphertext is transmitted to the receiver via the potentially insecure channel. Receiver runs the Rossler machine at his end and determines which final value corresponds to what initial value of the variable x, and is therefore able to decipher the ciphertext.

## Data Integrity Scheme

Individual characters of the plaintext message are first mapped to real numbers. In addition, the characters in the message are assigned positional weights. The weighted sum is computed. This weighted sum is used as the initial condition for the first component of the initial state vector in the Rossler machine. The Rossler machine is run with given state parameter and initial state vectors and number of time steps. The value at the end of the run is taken as the "message digest" corresponding to the message. This "message digest" is transmitted to the receiver via the potentially insecure channel. The receiver decrypts the ciphertext (if the message is encrypted) and subsequently runs the Rossler dynamics at his end and determines if the computed "message digest" matches the value given to him by the sender. If there is a match, then the message integrity is ensured, otherwise the message has been tampered with.

## Conclusions

The random-like dynamic behavior, high sensitivity to initial conditions, computational irreducibility of system trajectories, potentially large real phase space, potentially large space of real system parameters, and digital realization through non-integer arithmetic of continuous time chaotic systems are useful in the design of cryptography algorithms. The abstract Rossler machine has been described. Use of the Rossler machine in the design of algorithms for data confidentiality and integrity has been shown. Hardware implementation of the Rossler machine can be used in wireless sensor networks. The potential combinations of key components assuming an N-bit representation for floating point and integer numbers is $2^{7N}$, e.g., if N=16, we have $2^{102}$ potential combinations. This is beyond the reach of state-of-the-art supercomputers.